# Coherent deflection of Gaussian bunches colliding at crossing angle

A.A. Babaev

Tomsk Polytechnic University, Tomsk, Russia

Tomsk Polytechnic University, Lenin ave 30, Tomsk, 634050 Russia

2021/05/24

**Abstract.** The coherent deflection of elliptic bunch with the Gaussian charge distribution in the field of an opposite bunch with the Gaussian bunch distribution is considered. The generalization of Bassetti-Erskine formulae is derived taking into account the angle between momenta of bunches.

## 1 Introduction

At the collision of charged particles bunches in colliders one bunch is deflected by the electromagnetic field of the opposite bunch. This deflection is described by the coherent kick that is the average deflection of particles in the deflected bunch. The average procedure is performed taking into account the particles distribution in the bunch. This coherent kick leads to orbit shift relative to the design orbit. Here the 3D Gaussian distribution is considered. It is well-known in this case the coherent kick can be calculated using Bassetti-Erskine formulae and complex error functions (Faddeeva functions) [1,2,3]. However, these formulae were derived for counter-propagating bunches, whereas the real collision geometries (in particular, at LHC [4]) often involve non-zero crossing angles. At present, the problem is considered with numerical sumulations and simplifications reducing the problem to the known case of counter-parallel beams collision [5,6]. In particular, it is supposed the bunch shape in the corresponding coordinate frame remains Gaussian whereas this condition is broken due to relativistic effects.

The purpose of this note is to develop the general model including crossing angle where regular Bassetti-Erskine formulae is the particular case. In the note the procedure of deriving Bassetti-Erskine formulae [1,7,8] is extended. As in the regular derivation the collision is considered is the scheme where the first bunch (source bunch) generates the field deflecting the opposite bunch. The standard simplification on the small single-particle momentum kick relative to its initial momentum is used. The result is found for the important case of colliding protons of equal energies.

The note is structured as follows. In section 2 the geometry of the task and main definitions are introduced. In section 3 the single-particle kick is derived in the frame of the rest frame of source bumch. In section 4 the single-particle kick is transformed to the laboratory frame. In section 5 the coherent angular kick is calculated. In section 6 the generalization of Bassetti-Erskine formulae to the case of crossing angle is derived. Long mathematical procedures are taken out to appendices to avoid the mathematical cumbersome in the main text.

## 2 Geometry and definitions

Through the text the bunch 1 is the source of the field, bunch 2 is deflected by the field generated by the bunch 1. Correspondingly, the index 1 refers to quantities related to bunch 1 and the index 2 is related to the quantities characterizing the bunch 2 if another is not pointed out exactly. All particles in bunch 1 have the same initial momentum $\vec{p}_1$ and all particles in bunch 2 have the same initial momentum $\vec{p}_2$.

Here three laboratory frames are used (see in figure 1 and section 4): (i) the accelerator frame XYZ with the longitudinal axis Z, horizontal axis X, and vertical axis Y, (ii) the laboratory frame of bunch 1 $X_1Y_1Z_1$ where axis $Z_1$ is parallel to $\vec{p}_1$, (iii) the laboratory frame of bunch 2 $X_2Y_2Z_2$ where axis $Z_2$ is parallel to $\vec{p}_2$. It is assumed bunch dimensions are measured in its laboratory frame (i. e. the longitudinal size for bunch 1 is the size along $Z_1$, and the longitudinal size for bunch 2 is the size along $Z_2$ etc). Projections of momenta onto Z, $\vec{p}_{1z}$ and $\vec{p}_{2z}$, are counter-parallel. Projections of momentums onto X are positive. The angle between $\vec{p}_1$ and Z is $\theta$. The angle between $\vec{p}_1$ and $-\vec{p}_2$ (crossing angle) is $\alpha = 2\theta$, the angle between $-\vec{p}_2$ and Z is $\theta$.

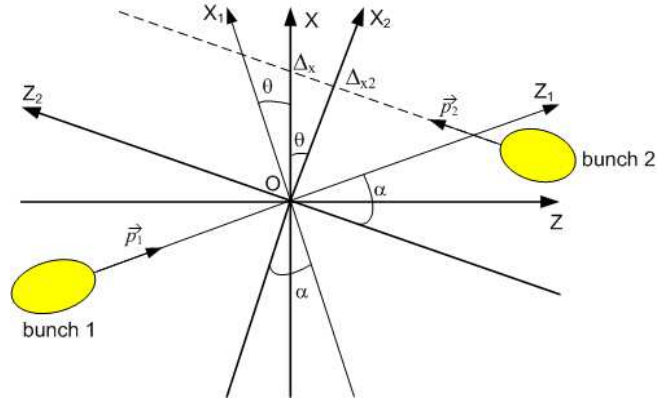

**Fig. 1.** Accelerator plane XZ and momenta $\vec{p}_1$, $\vec{p}_2$ in the accelerator frame. The beam separation in accelerator frame is $\Delta_x$. Additionally laboratory frames $X_1Z_1$ and $X_2Z_2$ are shown, see in section 4; $\Delta_{x2}$ is the projection of beam separation onto $X_2$, see in section 5.

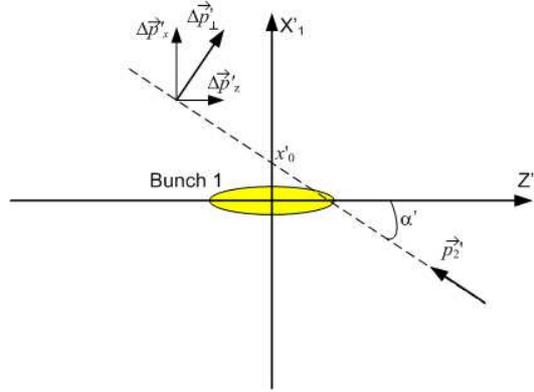

**Fig. 2.** Bunch 1 rest frame and trajectory of particle of bunch 2 in this frame (dashed line).

Also, in the note the beam separations $(\Delta_x, \Delta_y)$ are introduced. These separations define the distance from the XYZ frame origin to the point where bunch 2 orbit crosses the transverse plane XY at the assumption the orbit of bunch 1 crosses this plane at the origin. The separation $\Delta_x$ is counted in horizontal direction and it is shown in figure 1. The separation $\Delta_y$ is counted in vertical direction.

The interaction of colliding bunches is considered in the rest frame $X_1'Y_1'Z_1'$ of bunch 1. The longitudinal axis of the rest frame is $Z_1'$, the horizontal transverse axis is $X_1'$, the vertical axis is $Y_1'$. The last coincides with Y. Below primed quantities correspond to quantities in the rest frame of bunch 1. The angle between $Z_1'$ and $-\vec{p}_2'$ is $\alpha'$, $\alpha' \neq \alpha$ due to relativistic effects. For example, considering LHC conditions [4] where protons (the rest mass $\approx 1$ GeV) have momentum of the order 1-10 TeV/$c$ and crossing angle $\alpha$ is of the order 100 $\mu$rad in accelerator frame the transverse momentum of a proton is of the order 0.1-1 GeV that could be comparable to its rest mass. Therefore, relativistic effects in transverse motion could be important. The rest frame of bunch 1 and the trajectory of a particle of bunch 2 are shown in figure 2. As at the derivation of regular Bassetti-Erskine formulae it is assumed the betatron motion and radiation loss during the collision are not important. If particles distribution in bunch 1 is elliptic and main axes of ellipse coincide axes of frame (ii) $X_1Y_1Z_1$ then it remains elliptic in the rest frame but its dimensions differ from corresponding dimensions in the laboratory frame. In the rest frame the source bunch generates electric field only.

## 3 Single particle kick in bunch 1 rest frame

In the note the charge distribution in bunch 1 (in its rest frame, figure 2) is considered in 3D Gaussian form:

$$\rho_1(x',y',z') = \frac{eN_1}{(2\pi)^{3/2}\sigma_{1x}'\sigma_{1y}'\sigma_{1z}'} \exp\left[-\left(\frac{x'^2}{2\sigma_{1x}'^2} + \frac{y'^2}{2\sigma_{1y}'^2} + \frac{z'^2}{2\sigma_{1z}'^2}\right)\right].$$

In terms of normalized 1D Gaussian functions (A.1) this equation can be written as:

$$\rho_1(x',y',z') = eN_1 G(x';\sigma_{1x}')G(y';\sigma_{1y}')G(z';\sigma_{1z}'), \tag{1}$$

Here $N_1$ is the number of protons in bunch 1, $e$ is the proton charge. The normalization condition gives the total charge of bunch 1: $\int \rho_1 dx'dy'dz' = eN_1$ is the total charge of bunch 1. Electric field generated by distribution (1) was considered in [7]. Here it is used in terms of functions (A.2):

$$E'_{x1}(x', y', z') = -\frac{eN_1}{\sqrt{\pi}4\pi\epsilon_0}\frac{\partial}{\partial x'}\int_0^\infty Q(x', q; \sigma_{1x'})Q(y', q; \sigma_{1y'})Q(z', q; \sigma_{1z'})dq, \tag{2a}$$

$$E'_{y1}(x', y', z') = -\frac{eN_1}{\sqrt{\pi}4\pi\epsilon_0}\frac{\partial}{\partial y'}\int_0^\infty Q(x', q; \sigma_{1x'})Q(y', q; \sigma_{1y'})Q(z', q; \sigma_{1z'})dq, \tag{2b}$$

$$E'_{z1}(x', y', z') = -\frac{eN_1}{\sqrt{\pi}4\pi\epsilon_0}\frac{\partial}{\partial z'}\int_0^\infty Q(x', q; \sigma_{1x'})Q(y', q; \sigma_{1y'})Q(z', q; \sigma_{1z'})dq. \tag{2c}$$

These equations imply the origin of the bunch 1 rest frame coincides with the geometrical centre of bunch 1 and axes of this frame coincide with the axes of elliptical distribution (1).

Let consider the motion of a proton of bunch 2 in the rest frame of bunch 1 (see in figure 2). There are conventional assumptions (see, for example, in [9]): the electric field effectively influences the motion of particle only within relatively small duration of time, the transverse displacement of particle is negligible during this time, only momentum direction is influenced and momentum change $\Delta p'$ is small comparing to initial momentum $p'_2$. In the frame where only electric field exists the projection of $\Delta \vec{p'}$ onto horizontal $X'_1 Z'_1$ plane is orthogonal to trajectory line; this projection is $\Delta \vec{p'}_\perp = \Delta \vec{p'}_{x1} + \Delta \vec{p'}_{z1}$ where $\Delta p'_{x1} = e\int E'_{x1}dt'$ and $\Delta p'_{z1} = e\int E'_{z1}dt'$. From the geometry, $\Delta p'_{z1}/\Delta p'_{x1} = \tan\alpha'$. The vertical component of $\Delta \vec{p'}$ is $\Delta p'_{y1} = e\int E'_{y1}dt'$.

Let suppose the particle crosses $X'_1Y'_1$ plane with the offset $x'_0$ along $X'_1$ (see in figure 2) and offset $y'_0$ along $Y'_1$ at the time moment $t' = 0$. Therefore, the particle moves along the trajectory described by parametric equations $z' = v'_{2,z1}t'$, $x' = x'_0 + v'_{2,x1}t'$ where $v'_{2,x1}$, $v'_{2,z1}$ are projections of particle's velocity $\vec{v}'_2$ onto corresponding axes and $y' = y'_0 = \text{const}$. The trajectory can be considered like straight line if $\Delta p'_{x1} \leq 100$ MeV (i. e. the particle gets transverse momentum less than its initial transverse momentum, see the estimation in previous section).

The integration along the trajectory to find projections of $\Delta \vec{p'}$ results in

$$\Delta p'_{x1} = -\frac{e^2 N_1 \cos\alpha'}{4\pi\epsilon_0 |v'_{2,z1}|}\frac{\partial}{\partial x'_0}\int_0^\infty Q(x'_0\cos\alpha', q; S')Q(y', q; \sigma_{1y'})dq \tag{3a}$$

$$\Delta p'_{z1} = -\frac{e^2 N_1 \sin\alpha'}{4\pi\epsilon_0 |v'_{2,z1}|}\frac{\partial}{\partial x'_0}\int_0^\infty Q(x'_0\cos\alpha', q; S')Q(y', q; \sigma_{1y'})dq \tag{3b}$$

$$\Delta p'_{y1} = -\frac{e^2 N_1 \cos\alpha'}{4\pi\epsilon_0 |v'_{2,z1}|}\frac{\partial}{\partial y'}\int_0^\infty Q(x'_0\cos\alpha', q; S')Q(y', q; \sigma_{1y'})dq. \tag{3c}$$

where $S' = \sqrt{\sigma_{1x'}^2\cos^2\alpha' + \sigma_{1z'}^2\sin^2\alpha'}$ is the bunch 1 transverse cross-section relative to $\vec{p}'_2$. Details of calculations are summarized in appendix B.

## 4 Single-particle kick in laboratory frame

As mentioned above, three laboratory frames are used (see in figures 1 and 3): (i) the accelerator frame XYZ, (ii) the bunch 1 laboratory frame $X_1Y_1Z_1$, (iii) the bunch 2 laboratory frame $X_2Y_2Z_2$. The transitions between theese frames keep angles and lengths of vectors.

### 4.1 Lorentz transformations

Below bunches with the equal energy are considered; $\beta_1 = \beta_2 = \beta$, $\gamma_1 = \gamma_2 = \gamma$ where $\beta$ is the ratio of particle's velocity $v$ to speed of light $c$, $\gamma$ — relativistic factor. Following Lorentz transformations quantities in the rest frame of bunch 1 are related to the quantities in the laboratory frames by equations

$$\beta'_{2,z1} = -\frac{\beta(1+\cos\alpha)}{1+\beta^2\cos\alpha}, \quad \text{where } \beta'_{2,z1} = v'_{2,z1}/c$$

$$\beta'_{2,x1} = \frac{\beta\sin\alpha}{\gamma(1+\beta^2\cos\alpha)}, \quad \text{where } \beta'_{2,x1} = v'_{2,x1}/c$$

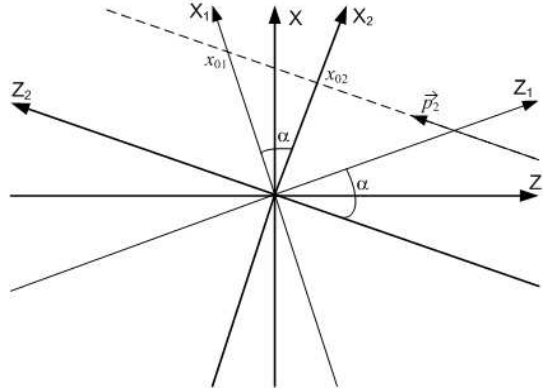

**Fig. 3.** Laboratory frames and motion of a particle of bunch 2.

$$\beta_2' = \sqrt{\beta'^2_{2,x1} + \beta'^2_{2,z1}} = \frac{\beta}{1+\beta^2\cos\alpha} f(\gamma,\alpha),$$

where

$$f(\gamma,\alpha) = \sqrt{\frac{\sin^2\alpha}{\gamma^2} + (1+\cos\alpha)^2}, \tag{4}$$

and $\gamma = 1/\sqrt{1-\beta^2}$. Corresponding equations for angles are:

$$\sin\alpha' = \frac{|\beta'_{2,x1}|}{\beta'_2} = \frac{\sin\alpha}{\gamma f(\gamma,\alpha)},$$

$$\cos\alpha' = \frac{|\beta'_{2,z1}|}{\beta'_2} = \frac{1+\cos\alpha}{f(\gamma,\alpha)}.$$

$$\tan\alpha' = \frac{|\beta'_{2,x1}|}{|\beta'_{2,z1}|} = \frac{\sin\alpha}{\gamma(1+\cos\alpha)}$$

As it was mentioned above, bunch dimensions are defined along axes of their laboratory frames; therefore, $\sigma_{1x'} = \sigma_{1x}$, $\sigma_{1y'} = \sigma_{1y}$, $\sigma_{1z'} = \gamma\sigma_{1z}$. In the laboratory frame the parameter $S'$ is transformed to

$$S = \frac{1}{f(\gamma,\alpha)}\sqrt{\sigma_{1x}^2(1+\cos\alpha)^2 + \sigma_{1z}^2\sin^2\alpha}. \tag{5}$$

The averaging in the laboratory frame over particles of bunch 2 (see in section 5) is performed at the fixed time moment. A particle with coordinates $(x_1, z_1)$ in horizontal plane of bunch 1 frame (ii) has coordinates $(x' = x_1, z' = \gamma z_1)$ at the corresponding time momentum in the rest frame of bunch 1. The particle crosses the axis $X_1'$ at the point $x_0' = x' + z'\tan\alpha'$. Therefore, the first parameter in functions $Q(x_0'\cos\alpha', q; S')$ in equations (3) is transformed as

$$x_0'\cos\alpha' = x'\cos\alpha' + z'\sin\alpha' = \frac{\xi}{f(\gamma,\alpha)}$$

and the derivative in (3a), (3b) in bunch 1 frame (ii) can be written as follows:

$$\frac{\partial}{\partial x_0'} \int_0^\infty Q(x_0'\cos\alpha', q; S') = (1+\cos\alpha)\frac{\partial}{\partial \xi}\int_0^\infty Q\left(\frac{\xi}{f(\gamma,\alpha)}, q; S\right)$$

where $\xi = x_1(1+\cos\alpha) + z_1\sin\alpha$. Coordinates $(x_1, z_1)$ must be taken at the same time moment for all particles in bunch 2. The $x_0'$ dependence on $(x_1, z_1)$ reveals the fact that particles moving along one trajectory sequentially in the laboratory frame do not move along one trajectory in the rest frame of bunch 1.

The vertical coordinate does not change: $y_1 = y'$. Vertical axes coincide in all laboratory frames and the subscript can be omitted: $y_1 = y_2 = y$.

Finally, projections of the momentum change vector $\Delta p$ in the laboratory frame of bunch 1 (ii) are: $\Delta p_{x1} = \Delta p'_{x1}$, $\Delta p_{z1} = \gamma\Delta p'_{z1}$, $\Delta p_y = \Delta p'_{y1}$ [2]. Using relations above equations (3) in this frame can be written in the form:

$$\Delta p_{x1} = -\frac{e^2 N_1(1+\cos\alpha)(1+\beta^2\cos\alpha)}{4\pi\epsilon_0 f(\gamma,\alpha)\beta c} \frac{\partial}{\partial \xi}\int_0^\infty Q\left(\frac{\xi}{f(\gamma,\alpha)}, q; S\right) Q(y, q; \sigma_{1y}) dq \tag{6a}$$

$$\Delta p_{z1} = -\frac{e^2 N_1 \sin\alpha (1 + \beta^2 \cos\alpha)}{4\pi\epsilon_0 f(\gamma,\alpha)\beta c} \frac{\partial}{\partial \xi} \int_0^\infty Q\left(\frac{\xi}{f(\gamma,\alpha)}, q; S\right) Q(y, q; \sigma_{1y}) dq \qquad (6b)$$

$$\Delta p_y = -\frac{e^2 N_1 (1 + \beta^2 \cos\alpha)}{4\pi\epsilon_0 f(\gamma,\alpha)\beta c} \frac{\partial}{\partial y} \int_0^\infty Q\left(\frac{\xi}{f(\gamma,\alpha)}, q; S\right) Q(y, q; \sigma_{1y}) dq. \qquad (6c)$$

where $f(\gamma,\alpha)$ and $S$ are defined by equations (4), (5), correspondingly.

The condition $\xi = \text{const}$ defines the line in horizontal plane of laboratory frame where particles in the bunch 2 get equal horizontal kicks $\Delta \vec{p}_{x1} + \Delta \vec{p}_{z1}$. The angle between this line and axis $Z_1$ is $-\arctan(\sin\alpha)/(1+\cos\alpha) = -\alpha/2$. In the geometry of figure 3 it means this line is parallel to axis Z. Also, the ratio $\Delta p_{z1}/\Delta p_{x1} = \tan(\alpha/2)$ and, therefore, the vector of horizontal kick is directed along axis X. In the frame XYZ $\xi = 2\cos(\alpha/2)x$ and $\partial/\partial\xi \propto \partial/\partial x$.

### 4.2 Single-particle deflection angle

The deflection angle for a particle from its initial motion direction in the horizontal plane is defined as $\phi_\perp = \Delta p_{x2}/p_2$ and the deflection angle in vertical direction is $\phi_y = \Delta p_y/p_2$ (at the condition $\Delta p \ll p_2$) where $p_2 = \beta\gamma mc$. The horizontal plane $X_2 Z_2$ of the frame of bunch 2 (iii) can be obtained from the plane $X_1 Z_1$ of the frame of bunch 1 (ii) by the clockwise rotation at the angle $\alpha$ and successive invertion of intermediate Z axis. The corresponding transformations for components of $\Delta\vec{p}$ and coordinates are

$$\begin{pmatrix} \Delta p_{x2} \\ \Delta p_{z2} \end{pmatrix} = \mathbf{M} \begin{pmatrix} \Delta p_{x1} \\ \Delta p_{z1} \end{pmatrix}, \quad \begin{pmatrix} x_2 \\ z_2 \end{pmatrix} = \mathbf{M} \begin{pmatrix} x_1 \\ z_1 \end{pmatrix}, \quad \text{where } \mathbf{M} = \begin{pmatrix} \cos\alpha & \sin\alpha \\ \sin\alpha & -\cos\alpha \end{pmatrix}$$

(frame $X_1 Y_1 Z_1$ is a right-handed frame). From this transformation $\Delta p_{x2} = \Delta p_{x1} \cos\alpha + \Delta p_{z1} \sin\alpha$. As $\mathbf{M}^{-1} = \mathbf{M}$ therefore $x_1 = x_2 \cos\alpha + z_2 \sin\alpha$, $z_1 = x_2 \sin\alpha - z_2 \cos\alpha$ and the parameter $\xi$ in the frame of bunch 2 (iii) is $\xi = x_2(1+\cos\alpha) + z_2 \sin\alpha$. The exact expression for $\Delta p_{x2}$ is obtained from equations (6a), (6b) and the transformation above:

$$\Delta p_{x2} = -\frac{e^2 N_1 (1 + \cos\alpha)(1 + \beta^2 \cos\alpha)}{4\pi\epsilon_0 f(\gamma,\alpha)\beta c} \frac{\partial}{\partial \xi} \int_0^\infty Q\left(\frac{\xi}{f(\gamma,\alpha)}, q; S\right) Q(y, q; \sigma_{1y}) dq \qquad (7)$$

The deflection angles in the horizontal plane and in the vertical direction are written using equations (7) and (6c), correspondingly:

$$\phi_\perp(x_2, y, z_2) = -\frac{N_1 r_0 (1 + \cos\alpha)(1 + \beta^2 \cos\alpha)}{f(\gamma,\alpha)\gamma\beta^2} \frac{\partial}{\partial \xi} \int_0^\infty Q\left(\frac{\xi}{f(\gamma,\alpha)}, q; S\right) Q(y, q; \sigma_{1y}) dq \qquad (8a)$$

$$\phi_y(x_2, y, z_2) = -\frac{N_1 r_0 (1 + \beta^2 \cos\alpha)}{f(\gamma,\alpha)\gamma\beta^2} \frac{\partial}{\partial y} \int_0^\infty Q\left(\frac{\xi}{f(\gamma,\alpha)}, q; S\right) Q(y, q; \sigma_{1y}) dq. \qquad (8b)$$

where $r_0 = e^2/(4\pi\epsilon_0 mc^2)$ is the classical radius of proton, $mc^2$ is the rest mass of proton.

## 5 Average deflection angle

In approximations used here the bunch distortion during the effective time of interaction is negligible. The density of particles in bunch 2 is considered 3D Gaussian at the moment when the bunch 2 centre crosses $X_2$ axis. The density in $X_2 Y_2 Z_2$ frame is:

$$n_2(x_2, y, z_2) = N_2 G(x_2 - \Delta_{x2}; \sigma_{2x}) G(y - \Delta_{y2}, \sigma_{2y}) G(z_2; \sigma_{2z}) \qquad (9)$$

where $\Delta_{x2}$, $\Delta_{y2}$ are introduced to define the bunch longitudinal axis (which is parallel to $Z_2$) transverse shift. These shifts are related to the nominal separations $\Delta_x$, $\Delta_y$ between bunches measured in the accelerator transverse plane XY as follows: $\Delta_{x2} = \Delta_x \cos\theta$, $\Delta_{y2} = \Delta_y$ (the longitudinal axis of bunch 1 concides with $Z_1$). Bunch dimensions transverse to its momentum are $\sigma_{2x}$ and $\sigma_{2y}$, the longitudinal beam size along $Z_2$ is defined by parameter $\sigma_{2z}$. Number of particles in bunch 2 is $N_2$. The normalization condition is $\int n_2 dx_2 dy_2 dz_2 = N_2$.

The average deflection angle is defined by the convolution of density (9) with the corresponding single-particle deflection angle (8a) or (8b) divided by normalization constant $N_2$. The calculations are performed using, in general, the procedure described in [8] and (A.14). See in appendix C for details. The convolution gives:

$$\langle\phi_\perp\rangle = -\frac{N_1 r_0 (1 + \beta^2 \cos\alpha)}{f(\gamma,\alpha)\gamma\beta^2} \frac{\partial}{\partial \Delta_{x2}} \int_0^\infty Q\left(\Delta_{x2} \frac{1 + \cos\alpha}{f(\gamma,\alpha)}, q; \Sigma_\perp\right) Q(\Delta_y, q; \Sigma_y) dq \qquad (10a)$$

$$\langle\phi_y\rangle = -\frac{N_1 r_0(1+\beta^2\cos\alpha)}{f(\gamma,\alpha)\gamma\beta^2}\frac{\partial}{\partial\Delta_y}\int_0^\infty Q\left(\Delta_{x2}\frac{1+\cos\alpha}{f(\gamma,\alpha)},q;\Sigma_\perp\right)Q(\Delta_y,q;\Sigma_y)dq \tag{10b}$$

where $\Sigma_y = \sqrt{\sigma_{1y}^2 + \sigma_{2y}^2}$ is the beam overlap width in the vertical direction of transverse accelerator plane XY and

$$\Sigma_\perp = \frac{1}{f(\gamma,\alpha)}\sqrt{(1+\cos\alpha)^2(\sigma_{1x}^2+\sigma_{2x}^2) + \sin^2\alpha(\sigma_{1z}^2+\sigma_{2z}^2)} \tag{11}$$

The parameter $\Sigma_\perp$ depends on beam energy; in general case, it can not be interpreted directly as beam overlap width. The beam overlap width in the accelerator frame in the direction X is:

$$\Sigma_x = \sqrt{\cos^2\theta(\sigma_{1x}^2+\sigma_{2x}^2) + \sin^2\theta(\sigma_{1z}^2+\sigma_{2z}^2)}$$

It is easy to show that

$$\Sigma_\perp = \frac{2\cos^2\theta}{f(\gamma,\alpha)}\Sigma_x$$

## 6 Generalization of Bassetti-Erskine formulae

For practical computations it is convinient to re-formulate equations (10) in terms of Faddeeva functions with complex argument

$$W(\tilde{z}) = e^{-\tilde{z}^2}\left(1 + \frac{2i}{\sqrt{\pi}}\int_0^{\tilde{z}} e^{\tilde{t}^2}d\tilde{t}\right)$$

In [1] it was shown that

$$-\frac{\partial}{\partial\zeta}\int_0^\infty Q(\zeta,q;\sigma_\zeta)Q(\eta,q;\sigma_\eta)dq = 2\pi\text{Im}(F(\zeta,\eta;\sigma_\zeta,\sigma_\eta) \tag{12a}$$

$$-\frac{\partial}{\partial\eta}\int_0^\infty Q(\zeta,q;\sigma_\zeta)Q(\eta,q;\sigma_\eta)dq = 2\pi\text{Re}(F(\zeta,\eta;\sigma_\zeta,\sigma_\eta) \tag{12b}$$

for planar orthogonal coordinate frame $(\zeta,\eta)$. Here

$$F(\zeta,\eta;\sigma_\zeta,\sigma_\eta) = \frac{1}{\sqrt{2\pi(\sigma_\zeta^2-\sigma_\eta^2)}}\left[W\left(\frac{\zeta+i\eta}{\sqrt{2(\sigma_\zeta^2-\sigma_\eta^2)}}\right) - \exp\left(-\frac{\zeta^2}{2\sigma_\zeta^2}-\frac{\eta^2}{2\sigma_\eta^2}\right)W\left(\frac{\frac{\sigma_\eta}{\sigma_\zeta}\zeta + i\frac{\sigma_\zeta}{\sigma_\eta}\eta}{\sqrt{2(\sigma_\zeta^2-\sigma_\eta^2)}}\right)\right]. \tag{13}$$

Comparing (12a) and (10a), (12b) and (10b) one can obtain:

$$\langle\phi_\perp\rangle = 2\pi\frac{N_1 r_0(1+\beta^2\cos\alpha)(1+\cos\alpha)}{f^2(\gamma,\alpha)\gamma\beta^2}\text{Im}\left[F\left(\Delta_{x2}\frac{1+\cos\alpha}{f(\gamma,\alpha)},\Delta_y;\Sigma_\perp,\Sigma_y\right)\right] \tag{14a}$$

$$\langle\phi_y\rangle = 2\pi\frac{N_1 r_0(1+\beta^2\cos\alpha)}{f(\gamma,\alpha)\gamma\beta^2}\text{Re}\left[F\left(\Delta_{x2}\frac{1+\cos\alpha}{f(\gamma,\alpha)},\Delta_y;\Sigma_\perp,\Sigma_y\right)\right] \tag{14b}$$

These equations are generalization of Bassetti-Erskine formulae to the case of arbitrary energies and crossing angles.

At small $\alpha$ one can obtain $f(\gamma,\alpha) \approx 2$, $\Delta_{x2} \approx \Delta_x$, $\Sigma_\perp \approx \tilde{\Sigma}_x = \sqrt{\sigma_{1x}^2+\sigma_{2x}^2+\theta^2(\sigma_{1z}^2+\sigma_{2z}^2)}$ and equations (14) are reduced to

$$\langle\phi_\perp\rangle = 2\pi\frac{N_1 r_0(1+\beta^2)}{2\gamma\beta^2}\text{Im}\left[F\left(\Delta_x,\Delta_y;\tilde{\Sigma}_x,\Sigma_y\right)\right]$$

$$\langle\phi_y\rangle = 2\pi\frac{N_1 r_0(1+\beta^2)}{2\gamma\beta^2}\text{Re}\left[F\left(\Delta_x,\Delta_y;\tilde{\Sigma}_x,\Sigma_y\right)\right]$$

In the ultrarelativistic limit ($\gamma \gg 1$, $\beta \approx 1$, $f(\gamma,\alpha) \approx 1+\cos\alpha$) at an arbitrary crossing angle equations (14) give:

$$\langle\phi_\perp\rangle = 2\pi\frac{N_1 r_0}{\gamma}\text{Im}\left[F\left(\Delta_{x2},\Delta_y;\tilde{\Sigma}_\perp,\Sigma_y\right)\right]$$

$$\langle\phi_y\rangle = 2\pi\frac{N_1 r_0}{\gamma}\text{Re}\left[F\left(\Delta_{x2},\Delta_y;\tilde{\Sigma}_\perp,\Sigma_y\right)\right],$$

where
$$\tilde{\Sigma}_\perp = \sqrt{\sigma_{1x}^2 + \sigma_{2x}^2 + \tan^2\theta(\sigma_{1z}^2 + \sigma_{2z}^2)}.$$

The ultrarelativistic limit and small $\alpha$ condition applied to equations (14) give:

$$\langle\phi_\perp\rangle = 2\pi\frac{N_1 r_0}{\gamma}\mathrm{Im}\left[F\left(\Delta_x, \Delta_y; \tilde{\Sigma}_x, \Sigma_y\right)\right]. \tag{15a}$$

$$\langle\phi_y\rangle = 2\pi\frac{N_1 r_0}{\gamma}\mathrm{Re}\left[F\left(\Delta_x, \Delta_y; \tilde{\Sigma}_x, \Sigma_y\right)\right], \tag{15b}$$

At $\alpha = 0$ equations (15) match regular Bassetti-Erskine formulae exactly.

The final conclusion of this section is: equations (14) represent the generalization of Bassetti-Erskine formulae to the case of arbitrary crossing angle and arbitrary energies. At the condition of modern colliders which operates with ultra-relativistic hadron beams equations (14) can be simplified to equations (15).

## 7 Final remarks

In this note the problem of average angular kick at collision of Gaussian bunches at arbitrary crossing angle is considered. The corresponding generalization of Bassetti-Erskine formulae is derived in the form (14). The result can be used in accelerator physics applications to model the beam dynamics in colliders [3]. In particular, the angular kick like (14) is used for computations of design orbit shift arising due to interaction of colliding bunches [3,13]. The design orbit perturbations of this kind are important at specific experiments like beam separation scans performed to precise luminosity calibration in colliders [10,14]. At beam separation scan beams are moved in transverse accelerator plane; i.e. $\Delta_x$, $\Delta_y$ are varied and, correspondingly, the angular deflection varies during the experiment.

There are few benefits of the work. First, this is the general result and regular Bassetti-Erskine formulae are its particular case. This general result can be used not only in analysis of ultra-relativistic beam collisions. Also, limits of regular Bassetti-Erskine formulae now can be understood clearly. Second, the task on the deflection in electric field generated by Gaussian-like distributed charge that is partly considered in section 3 can be applied to another tasks of physics abroad collider applications.

Third, the scheme developed here can be applied at any integrable particles distribution. The task is considered as the regular elastic scattering task. Initially the single particle kick is considered in the rest frame of source bunch and then the averaging is performed in the laboratory frame. This division of procedures allows to avoid the problem of bunch shape distortion at the Lorentz transformation from the laboratory frame to the rest frame of source bunch.

The presented approach can be easily expanded to the case when colliding particles of equal energies ($\gamma_1 = \gamma_2 = \gamma$) carry the charge $Z_1 e$ in the source beam and $Z_2 e$ in the opposite one. The result is the factor $Z_1 Z_2 r_0$ instead of $r_0$ in all equations.

Finally, one can consider the simple and important case when beams are accelerated by the same voltage up to ultra-relativistic energies; so, they get the energy $Z_1 E_0$ and $Z_2 E_0$ where $E_0$ is the so-called proton-equivalent energy. In this case one can substitute $\gamma_0 = E_0/(mc^2)$ and $Z_1 r_0$ in equations (15) instead of $\gamma$ and $r_0$, correspondingly. The careful consideration of general problem when colliding bunches have different energies is abroad of this note.

## A Functions $G(x;\sigma)$ and $Q(x,q;\sigma)$

Definitions of functions $G(x;\sigma)$ and $Q(x,q;\sigma)$:

$$G(x;\sigma) = \frac{1}{\sqrt{2\pi}\sigma}\exp\left(-\frac{x^2}{2\sigma^2}\right) \tag{A.1}$$

$$Q(x,q;\sigma) = \frac{1}{\sqrt{2\sigma^2+q}}\exp\left(-\frac{x^2}{2\sigma^2+q}\right) \tag{A.2}$$

Simple transformations:
$$Q(x,q;\sigma) = \sqrt{\pi}G(x;a(\sigma,q)), \quad \text{where } 2a^2 = 2\sigma^2 + q \tag{A.3}$$

$$G(bx - A;\sigma) = \frac{1}{b}G(x - A';\sigma'), \quad \text{where } b > 0, \quad A' = \frac{A}{b} \text{ and } \sigma' = \frac{\sigma}{b} \tag{A.4}$$

$$Q(bx - A, q;\sigma) = \frac{1}{b}Q(x - A', q';\sigma'), \quad \text{where } b > 0, \quad A' = \frac{A}{b}, \quad q' = \frac{q}{b^2}, \text{ and } \sigma' = \frac{\sigma}{b} \tag{A.5}$$

$$Q(x, aq; \sigma) = \frac{1}{\sqrt{a}} Q(x/\sqrt{a}, q; \sigma/\sqrt{a}), \quad a > 0 \tag{A.6}$$

Derivative of $G(x; \sigma)$:

$$\frac{dG(x; \sigma)}{dx} = -\frac{x}{\sigma^2} G(x; \sigma) \tag{A.7}$$

Derivative of $Q(x, q; \sigma)$ can be calculated from (A.3), (A.7) or directly from definition (A.2):

$$\frac{dQ(x, q; \sigma)}{dx} = -\frac{2x}{2\sigma^2 + q} Q(x, q; \sigma) \tag{A.8}$$

Normalization condition for $G(x; \sigma)$:

$$\int_{-\infty}^{\infty} G(x; \sigma) dx = 1 \tag{A.9}$$

Overlap integral is calculated directly from the definition (A.1):

$$\int_{-\infty}^{\infty} G(x - A; \sigma_1) G(x; \sigma_2) dx = G\left(A; \sqrt{\sigma_1^2 + \sigma_2^2}\right) \tag{A.10}$$

"Scaled overlap" integral for functions $Q$ is calculated using (A.5), (A.3), (A.10):

$$\int_{-\infty}^{\infty} Q(bx - A, q; \sigma_1) Q(x, q; \sigma_2) dx = \sqrt{\pi} Q\left(A, (1 + b^2)q; \sqrt{\sigma_1^2 + \sigma_2^2 b^2}\right) \tag{A.11}$$

Another integrals:

$$\int_{-\infty}^{\infty} x G(x - A; \sigma_1) G(x; \sigma_2) dx = \frac{\sigma_2^2 A}{\sigma_1^2 + \sigma_2^2} G\left(A; \sqrt{\sigma_1^2 + \sigma_2^2}\right) \tag{A.12}$$

The integration in (A.12) is performed using the definition (A.1), integration by parts, and (A.10).

$$\int_{-\infty}^{\infty} x Q(bx - A, q; \sigma_1) Q(x, q; \sigma_2) dx = \frac{\sqrt{\pi}(2\sigma_2^2 + q) Ab}{2(\sigma_1^2 + \sigma_2^2 b^2) + (1 + b^2)q} Q\left(A, (1 + b^2)q; \sqrt{\sigma_1^2 + \sigma_2^2 b^2}\right) \tag{A.13}$$

The integration in (A.13) is performed using (A.5), (A.3), (A.12) and there $b > 0$.

$$\int_{-\infty}^{\infty} G(x - A; \sigma_1) Q(bx, q; \sigma_2) dx = Q\left(bA, q; \sqrt{\sigma_1^2 b^2 + \sigma_2^2}\right), \quad b > 0 \tag{A.14}$$

The integration in (A.14) is performed using (A.3), (A.4), (A.10).

## B Integration along the trajectory in rest frame of bunch 1

Let consider the motion of particle in electric field (2) along the trajectory described by parametric equations $z' = v'_{2,z1} t'$, $x' = x'_0 + v'_{2,x1} t'$, and $y' = y'_0 = $ const.

First, let derive the formulae (3b) from the base equation $\Delta p'_{z1} = e \int E'_{z1} dt'$. Derivatives in (2) do not commute with the integration over $t'$ due to the coupling between $x'$ and $z'$. The derivative in (2c) is calculated directly using (A.8):

$$E'_{z1}(x', y', z') = \frac{2eN_1 z'}{\sqrt{\pi} 4\pi\epsilon_0} \int_0^{\infty} Q(x', q; \sigma_{1x'}) Q(y', q; \sigma_{1y'}) \frac{Q(z', q; \sigma_{1z'})}{2\sigma_{1z'}^2 + q} dq.$$

Then, trajectory equations are substituted in the integral which defines $\Delta p'_{z1}$:

$$\Delta p'_{z1} = \frac{2e^2 N_1 v'_{2,z1}}{\sqrt{\pi} 4\pi\epsilon_0} \int_{-\infty}^{\infty} \int_0^{\infty} Q(x'_0 + v'_{2,x1} t', q; \sigma_{1x'}) Q(y', q; \sigma_{1y'}) \frac{t' Q(v'_{2,z1} t', q; \sigma_{1z'})}{2\sigma_{1z'}^2 + q} dq dt'.$$

After that the order of integrals can be changed:

$$\Delta p'_{z1} = \frac{2e^2 N_1 v'_{2,z1}}{\sqrt{\pi} 4\pi\epsilon_0} \int_0^{\infty} \frac{Q(y', q; \sigma_{1y'})}{2\sigma_{1z'}^2 + q} \left[\int_{-\infty}^{\infty} t' Q(x'_0 + v'_{2,x1} t', q; \sigma_{1x'}) Q(v'_{2,z1} t', q; \sigma_{1z'}) dt'\right] dq.$$

The integral in squared brackets is reduced to (A.13) by using the substitution $t' \to v'_{2,z1}t'$. Hence, the integration in squared brackets gives:

$$\int_{-\infty}^{\infty} t'Q(x'_0 + v'_{2,x1}t', q; \sigma_{1x'})Q(v'_{2,z1}t', q; \sigma_{1z'})dt' =$$
$$-\frac{1}{v'^2_{2,z1}} \frac{\sqrt{\pi}(2\sigma^2_{1z'} + q)x'_0 \tan \alpha'}{2(\sigma^2_{1x'} + \sigma^2_{1z'} \tan^2 \alpha') + (1+\tan^2\alpha')q} Q\left(-x'_0, (1+\tan^2\alpha')q; \sqrt{\sigma^2_{1x'} + \sigma^2_{1z'}\tan^2\alpha'}\right).$$

where from the geometry $v'_{2,x1}/v'_{2,z1} = \tan\alpha'$. From the geometry $v'_{2,z1} < 0$ and

$$\Delta p'_{z1} = \frac{2e^2 N_1 x'_0 \tan\alpha'}{4\pi\epsilon_0 |v'_{2,z1}|} \int_0^\infty \frac{Q(y', q; \sigma_{1y'})Q\left(x'_0, q/\cos^2\alpha'; \sqrt{\sigma^2_{1x'} + \sigma^2_{1z'}\tan^2\alpha'}\right)}{2(\sigma^2_{1x'} + \sigma^2_{1z'}\tan^2\alpha') + q/\cos^2\alpha'} dq,$$

After that (A.8) is used:

$$\Delta p'_{z1} = -\frac{e^2 N_1 \tan\alpha'}{4\pi\epsilon_0 |v'_{2,z1}|} \frac{\partial}{\partial x'_0} \int_0^\infty Q\left(x'_0, q/\cos^2\alpha'; \sqrt{\sigma^2_{1x'} + \sigma^2_{1z'}\tan^2\alpha'}\right) Q(y', q; \sigma_{1y'})dq$$

And, finally, (A.6) is used:

$$\Delta p'_{z1} = -\frac{e^2 N_1 \sin\alpha'}{4\pi\epsilon_0 |v'_{2,z1}|} \frac{\partial}{\partial x'_0} \int_0^\infty Q\left(x'_0 \cos\alpha', q; \sqrt{\sigma^2_{1x'}\cos^2\alpha' + \sigma^2_{1z'}\sin^2\alpha'}\right) Q(y', q; \sigma_{1y'})dq \qquad (B.1)$$

The equation (B.1) is the equation (3b). The equation (3a) can be obtained directly from (B.1) using geometrical relation $\Delta p'_{z1}/\Delta p'_{x1} = \tan\alpha'$.

Let calculate the integral in $\Delta p'_{y1} = e \int E'_{y1} dt'$ where $E'_{y1}$ is defined by equation (2b) and trajectory equations are written above:

$$\Delta p'_{y1} = -\frac{e^2 N_1}{\sqrt{\pi}4\pi\epsilon_0} \frac{\partial}{\partial y'} \int_0^\infty Q(y', q; \sigma_{1y'}) \left[\int_{-\infty}^{-\infty} Q(x'_0 + v'_{2,x1}t', q; \sigma_{1x'})Q(v'_{2,z1}t', q; \sigma_{1z'})dt\right] dq.$$

Here it is supposed the derivation is calculated first and after that the vertical coordinate is fixed corresponding to $y' = y'_0 = $ const. The integral in squared brackets is calculated using the substitution $t' \to v'_{2,z1}t'$, (A.11), (A.6):

$$\int_{-\infty}^{-\infty} Q(x'_0 + v'_{2,x1}t', q; \sigma_{1x'})Q(v'_{2,z1}t', q; \sigma_{1z'})dt = -\frac{\sqrt{\pi}}{|v'_{2,z1}|} Q\left(-x'_0, (1+\tan^2\alpha')q; \sqrt{\sigma^2_{1x'} + \sigma^2_{1z'}\tan^2\alpha'}\right)$$

and, therefore,

$$\Delta p'_{y1} = -\frac{e^2 N_1}{4\pi\epsilon_0 |v'_{2,z1}|} \frac{\partial}{\partial y'} \int_0^\infty Q\left(x'_0, q/\cos^2\alpha'; \sqrt{\sigma^2_{1x'} + \sigma^2_{1z'}\tan^2\alpha'}\right) Q(y', q; \sigma_{1y'})dq.$$

And, finally, (A.6) is used:

$$\Delta p'_{y1} = -\frac{e^2 N_1 \cos\alpha'}{4\pi\epsilon_0 |v'_{2,z1}|} \frac{\partial}{\partial y'} \int_0^\infty Q(x'_0 \cos\alpha', q; \sqrt{\sigma^2_{1x'}\sin^2\alpha' + \sigma^2_{1z'}\sin^2\alpha'})Q(y', q; \sigma_{1y'})dq. \qquad (B.2)$$

and equation (3c) is derived.

It worth to be mentioned the argument $x'_0 \cos\alpha' = \xi'$ in (B.1), (B.2) is the distance from bunch 1 centre to particle's trajectory. In terms of $\xi'$ equations (B.1), (B.2) can be written in the form:

$$\Delta p'_{z1} = -\frac{e^2 N_1 \sin\alpha' \cos\alpha'}{4\pi\epsilon_0 |v'_{2,z1}|} \frac{\partial}{\partial \xi'} \int_0^\infty Q\left(\xi', q; \sqrt{\sigma^2_{1x'}\cos^2\alpha' + \sigma^2_{1z'}\sin^2\alpha'}\right) Q(y', q; \sigma_{1y'})dq$$

$$\Delta p'_{y1} = -\frac{e^2 N_1 \cos\alpha'}{4\pi\epsilon_0 |v'_{2,z1}|} \frac{\partial}{\partial y'} \int_0^\infty Q(\xi', q; \sqrt{\sigma^2_{1x'}\sin^2\alpha' + \sigma^2_{1z'}\sin^2\alpha'})Q(y', q; \sigma_{1y'})dq.$$

## C Averaging procedure

The average deflection angle in the vertical direction, $\langle\phi_y\rangle$ is defined by the integral

$$\langle\phi_y\rangle = \frac{1}{N_2}\int n_2(x_2,y,z_2)\phi_y(x_2,y,z_2)dx_2dydz_2$$

where $n_2(x_2,y,z_2)$ is defined by (9), $\phi_y(x_2,y,z_2)$ is defined by (8b). This integral is expanded in the form

$$\langle\phi_y\rangle = -C\int_{-\infty}^{\infty}\int_{-\infty}^{\infty}\int_{-\infty}^{\infty} G(x_2-\Delta_{x2};\sigma_{2x})G(y-\Delta_y,\sigma_{2y})G(z_2;\sigma_{2z})$$
$$\times \frac{\partial}{\partial y}\left[\int_0^{\infty} Q\left(\frac{x_2(1+\cos\alpha)+z_2\sin\alpha}{f(\gamma,\alpha)},q;S\right)Q(y,q;\sigma_{1y})dq\right]dx_2dydz_2 \quad (C.1)$$

where the definition $\xi = x_2(1+\cos\alpha)+z_2\sin\alpha$ is used, $C = N_1 r_0(1+\beta^2\cos\alpha)/(f(\gamma,\alpha)\gamma\beta^2)$.

The integration over $z_2$ commutes with another operations in (C.1). The use of (A.14) leads to

$$\int_{-\infty}^{\infty} G(z_2;\sigma_{2z})Q\left(\frac{x_2(1+\cos\alpha)+z_2\sin\alpha}{f(\gamma,\alpha)},q;S\right)dz_2 = Q\left(\frac{x_2(1+\cos\alpha)}{f(\gamma,\alpha)},q;\tilde{S}\right)$$

where

$$\tilde{S} = \sqrt{\left(\frac{\sigma_{2z}\sin\alpha}{f(\gamma,\alpha)}\right)^2 + S^2} \quad (C.2)$$

Hence, (C.1) is reduced to

$$\langle\phi_y\rangle = -C\int_{-\infty}^{\infty}\int_{-\infty}^{\infty} G(x_2-\Delta_{x2};\sigma_{2x})G(y-\Delta_y,\sigma_{2y})$$
$$\times \frac{\partial}{\partial y}\left[\int_0^{\infty} Q\left(\frac{x_2(1+\cos\alpha)}{f(\gamma,\alpha)},q;\tilde{S}\right)Q(y,q;\sigma_{1y})dq\right]dx_2dy \quad (C.3)$$

The derivation is transferred from the function $Q$ to the function $G$ using the integratiion by parts and zero limits of (A.1) at infinity. The result is:

$$\langle\phi_y\rangle = C\int_{-\infty}^{\infty}\int_{-\infty}^{\infty} G(x_2-\Delta_{x2};\sigma_{2x})\frac{\partial G(y-\Delta_y;\sigma_{2y})}{\partial y}\left[\int_0^{\infty} Q\left(\frac{x_2(1+\cos\alpha)}{f(\gamma,\alpha)},q;\tilde{S}\right)Q(y,q;\sigma_{1y})dq\right]dx_2dy$$

After that the property

$$\frac{\partial G(y-\Delta_y;\sigma_{2y})}{\partial y} = -\frac{\partial G(y-\Delta_y;\sigma_{2y})}{\partial\Delta_y}$$

is used with the result:

$$\langle\phi_y\rangle = -C\frac{\partial}{\partial\Delta_y}\int_{-\infty}^{\infty}\int_{-\infty}^{\infty} G(x_2-\Delta_{x2};\sigma_{2x})G(y-\Delta_y,\sigma_{2y})\left[\int_0^{\infty} Q\left(\frac{x_2(1+\cos\alpha)}{f(\gamma,\alpha)},q;\tilde{S}\right)Q(y,q;\sigma_{1y})dq\right]dx_2dy$$

Now variables can be uncoupled:

$$\langle\phi_y\rangle = -C\frac{\partial}{\partial\Delta_y}\int_0^{\infty}\left[\int_{-\infty}^{\infty} G(x_2-\Delta_{x2};\sigma_{2x})Q\left(\frac{x_2(1+\cos\alpha)}{f(\gamma,\alpha)},q;\tilde{S}\right)dx_2 \int_{-\infty}^{\infty} G(y-\Delta_y,\sigma_{2y})Q(y,q;\sigma_{1y})dy\right]dq$$

and (A.14) can be used:

$$\langle\phi_y\rangle = -C\frac{\partial}{\partial\Delta_y}\int_0^{\infty} Q\left(\Delta_{x2}\frac{1+\cos\alpha}{f(\gamma,\alpha)},q;\Sigma_\perp\right)Q(\Delta_y,q;\Sigma_y)dq$$

where $\Sigma_y = \sqrt{\sigma_{1y}^2 + \sigma_{2y}^2}$ and

$$\Sigma_\perp = \sqrt{\tilde{S}^2 + \left(\frac{(1+\cos\alpha)\sigma_{2x}}{f(\gamma,\alpha)}\right)^2}$$

The last expression can be written in the form:

$$\Sigma_\perp = \frac{1}{f(\gamma,\alpha)}\sqrt{(1+\cos\alpha)^2(\sigma_{1x}^2+\sigma_{2x}^2)+\sin^2\alpha(\sigma_{1z}^2+\sigma_{2z}^2))} \tag{C.4}$$

where (C.2), (5) have been used. So, the equation (10b) is derived.

Calculations for the averaged angle $\langle\phi_\perp\rangle$ (10b) are performed starting from the integral

$$\langle\phi_\perp\rangle = \frac{1}{N_2}\int n_2(x_2,y,z_2)\phi_\perp(x_2,y,z_2)dx_2dydz_2$$

where $\phi_\perp(x_2,y,z_2)$ is defined by (8a). The expanded form of this integral is

$$\langle\phi_\perp\rangle = -C(1+\cos\alpha)\int_{-\infty}^{\infty}\int_{-\infty}^{\infty}\int_{-\infty}^{\infty} G(x_2-\Delta_{x2};\sigma_{2x})G(y-\Delta_y,\sigma_{2y})G(z_2;\sigma_{2z})$$
$$\times \frac{\partial}{\partial\xi}\left[\int_0^\infty Q\left(\frac{\xi}{f(\gamma,\alpha)},q;S\right)Q(y,q;\sigma_{1y})dq\right]dx_2dydz_2 \tag{C.5}$$

It is convinient to consider the integral (C.5) in frame XYZ. Indeed, in section 4.1 is was shown $\xi = 2x\cos\theta$ and, correspondingly, $\partial/\partial\xi = 1/(2\cos\theta)\times\partial/\partial x$ where $\theta = \alpha/2$. The transfer from frame XYZ to frame $X_2Y_2Z_2$ follows the matrix equation

$$\begin{pmatrix}x_2\\z_2\end{pmatrix} = \begin{pmatrix}\cos\theta & \sin\theta\\ \sin\theta & -\cos\theta\end{pmatrix}\begin{pmatrix}x\\z\end{pmatrix}$$

Hence, equation (C.5) is rewritten in the form:

$$\langle\phi_\perp\rangle = -C\cos\theta\int_{-\infty}^{\infty}\int_{-\infty}^{\infty}\left[\int_{-\infty}^{\infty}G(x\cos\theta+z\sin\theta-\Delta_{x2};\sigma_{2x})G(x\sin\theta-z\cos\theta;\sigma_{2z})\right]dz$$
$$\times G(y-\Delta_y,\sigma_{2y})\frac{\partial}{\partial x}\left[\int_0^\infty Q\left(x\frac{2\cos\theta}{f(\gamma,\alpha)},q;S\right)Q(y,q;\sigma_{1y})dq\right]dxdy$$

The integral over $z$ in squared brackets is calculated using (A.4), (A.10) with the result

$$\langle\phi_\perp\rangle = -C\cos\theta\int_{-\infty}^{\infty}\int_{-\infty}^{\infty}G(x-\Delta_{x2}\cos\theta;S_2)G(y-\Delta_y,\sigma_{2y})$$
$$\times \frac{\partial}{\partial x}\left[\int_0^\infty Q\left(x\frac{2\cos\theta}{f(\gamma,\alpha)},q;S\right)Q(y,q;\sigma_{1y})dq\right]dxdy \tag{C.6}$$

where

$$S_2 = \sqrt{\sigma_{2x}^2\cos^2\theta+\sigma_{2z}^2\cos^2\theta} \tag{C.7}$$

Equations (C.6) and (C.3) have the same structure; integration in (C.6) can be performed with using the same procedure. The result is

$$\langle\phi_\perp\rangle = -C\frac{\partial}{\partial\Delta_{x2}}\int_0^\infty Q\left(\Delta_{x2}\frac{1+\cos\alpha}{f(\gamma,\alpha)},q;\Sigma_\perp\right)Q(\Delta_y,q;\Sigma_y)dq$$

where

$$\Sigma_\perp = \sqrt{\left(\frac{2\cos\theta}{f(\gamma,\alpha)}\right)^2 S_2^2 + S^2}$$

It could be shown using (C.7), (5) this expression for $S_\perp$ is the same (C.4). So, the equation (10a) is derived.